\documentclass[conference,a4paper]{IEEEtran}
\IEEEoverridecommandlockouts
\usepackage{cite}
\usepackage{amsmath,amssymb,amsfonts}
\usepackage{algorithmic}
\usepackage{graphicx}
\usepackage{textcomp}
\usepackage{xcolor}
\usepackage{bm,bbm}
\usepackage{lipsum}
\usepackage{subfigure}

\def\BibTeX{{\rm B\kern-.05em{\sc i\kern-.025em b}\kern-.08em
    T\kern-.1667em\lower.7ex\hbox{E}\kern-.125emX}}

\newcommand{\E}{\mathbb{E}}

\newtheorem{Theorem}{Theorem}

\begin{document}

\title{Average Age of Information Penalty of Short-Packet Communications with Packet Management}

%\author{\IEEEauthorblockN{Zhifeng Tang, Nan Yang, Xiangyun Zhou}
%\IEEEauthorblockA{School of Engineering, Australian National University, Canberra, ACT 2600, Australia}
%\IEEEauthorblockA{Email: zhifeng.tang@anu.edu.au, nan.yang@anu.edu.au, xiangyun.zhou@anu.edu.au}}

\author{\IEEEauthorblockN{Zhifeng Tang$^{\ast}$, Nan Yang$^{\ast}$, Xiangyun Zhou$^{\ast}$, and Jemin Lee$^{\dagger}$}
\IEEEauthorblockA{$^{\ast}$School of Engineering, Australian National University, Canberra, ACT 2600, Australia}
\IEEEauthorblockA{$^{\dagger}$Department of Electrical and Computer Engineering, Sungkyunkwan University, Suwon 16419, South Korea}
\IEEEauthorblockA{Email: zhifeng.tang@anu.edu.au, nan.yang@anu.edu.au, xiangyun.zhou@anu.edu.au, jemin.lee@skku.edu}}

\maketitle

\begin{abstract}
In this paper, we analyze the non-linear age of information (AoI) performance in a point-to-point short packet communication system, where a transmitter generates packets based on status updates and transmits the packets to a receiver. Specifically, we investigate three packet management strategies, namely, the non-preemption with no buffer strategy, the non-preemption with one buffer strategy, and the preemption strategy. To characterize the level of the receiver's dissatisfaction on outdated data, we adopt a generalized $\alpha$-$\beta$ AoI penalty function into the analysis and derive closed-form expressions for the average AoI penalty achieved by the three packet management strategies. Simulation results are used to corroborate our analysis and explicitly evaluate the impact of various system parameters, such as the coding rate and status update generation rate, on the AoI performance. Additionally, we find that the value of $\alpha$ reflects the system transmission reliability. 
%show that the receiver requires higher transmission reliability when it has a relatively low tolerance to the outdated data, but a lower reliability when it has a relatively high tolerance to the outdated data.
\end{abstract}


\IEEEpeerreviewmaketitle

\section{Introduction}

Effective communication paradigms to support the connectivity in real-time applications, such as intelligent transport systems and factory automation, have recently attracted a wide range of interests. In these applications, timely status updates are pivotal requirements for realizing accurate monitoring and control \cite{Li2019}. To meet such requirements, short packet communications has been recognized as a promising technique to reduce transmission latency %in real-time applications
\cite{Sun2018,Huang2019,Li2021}. Moreover, the age of information (AoI) has been introduced as a new and effective performance metric to characterize the freshness of transmitted status information \cite{Kaul2011}. Specifically, the AoI is defined as the elapsed time since the last successfully received status was generated by the transmitter, which is a time metric capturing both latency and freshness of transmitted status information. Since being introduced \cite{Kaul2011}, the AoI has reaped a wealth of attention. For example, some studies have analyzed the AoI performance, e.g., \cite{Tang2020,Wang2020IoT,Tang2021GC}, while some other studies have designed transmission policies to improve the AoI performance, e.g., \cite{Wang2019a,Kadota2016,Hsu2018,Sun2019A}.

%Starting from analyzing the AoI performance in \cite{Tang2020,Wang2020IoT,Tang2021GC}, some transmission policies were designed to effectively improve the AoI performance \cite{Wang2019a,Kadota2016,Hsu2018,Sun2019A}.

Recently, the non-linear AoI performance has been examined as an important extension to the average AoI performance \cite{Kosta2020,Sun2016Fresh,Zheng2019NL,Klugel2019,tang2021whittle,Hu2022Penalty}, due to its benefits of characterizing how the receiver's dissatisfaction on the freshness of its received information depends on data staleness. In \cite{Kosta2020}, the average cost of AoI was derived for three sample functions in an M/M/1 queue model with a first-come-first-served (FCFS) queuing policy. Moreover, \cite{Sun2016Fresh} proposed an exponential age penalty function to characterize the receiver's dissatisfaction with the degree of data staleness. Considering an energy harvesting system, \cite{Zheng2019NL} evaluated the average exponential AoI with FCFS and last-come-first-served (LCFS) queuing policies. In \cite{Klugel2019}, a threshold based scheduling policy was proposed to optimize a general average cost of AoI in network control systems. Considering a general non-decreasing function of the AoI, \cite{tang2021whittle} designed the Whittle index based scheduling policy. Very recently, \cite{Hu2022Penalty} proposed an $\alpha$-$\beta$ AoI penalty function to characterize different nonlinear forms of AoI penalty and investigated the average $\alpha$-$\beta$ AoI penalty of a wireless-powered communication system.

Motivated by the benefits of short packet communications for latency reduction, the AoI performance of short packet communications was analyzed by \cite{devassy2018delay,Devassy2019,Basnay2021,WangGC2019} to examine the impact of short packets on information freshness. Specifically, \cite{devassy2018delay} investigated the impact of the packet blocklength on the delay and peak AoI in a point-to-point communication system. Considering the same system, \cite{Devassy2019}
%extended \cite{devassy2018delay} to
analyzed the probability that the peak-age violation exceeds a desired threshold. Focusing on a decode-and-forward relay system, \cite{Basnay2021} estimated the impact of the packet generation rate, packet blocklength, and blocklength allocation factor on the average AoI. Furthermore, \cite{WangGC2019} studied the optimal packet blocklength of non-preemption and preemption strategies for minimizing the average AoI. Although the aforementioned studies have evaluated on the linear AoI performance of short packet communication systems with different packet management strategies, the non-linear AoI performance of short packet communication systems remains untouched, which motivates this work.

In this paper, we analytically assess the impact of three packet management strategies on the non-linear AoI performance of a point-to-point short packet communication system. In this system, the transmitter generates status updates and transmits them to the receiver via an unreliable channel. We examine three packet management strategies, namely, the non-preemption with no buffer (NPNB) strategy, the non-preemption with one buffer (NPOB) strategy, and the preemption strategy. To characterize the level of the receiver's dissatisfaction on outdated status update, we derive new closed-form expressions for the average AoI penalty achieved by three packet management strategies, where the generalized $\alpha$-$\beta$ AoI penalty function \cite{Hu2022Penalty} is adopted. Aided by simulations, we demonstrate the accuracy of our analytical results and find the relationship between the requirement of the transmission reliability and the toleration on the outdated date. %and discuss the impacts of various parameters on the average AoI penalty. %with three packet management strategies.
%We find that the NPOB strategy achieves a lower average AoI penalty than the NPNB strategy for a small status update generation rate, but the NPNB strategy achieves a lower average AoI penalty for a large status update generation rate. We also find that high transmission reliability is required when the system has a low tolerance to outdated status update.

\section{System Model and Average AoI Penalty}\label{Sec:System_Model}

\begin{figure}[t]
\centering
\includegraphics[width=0.85\columnwidth]{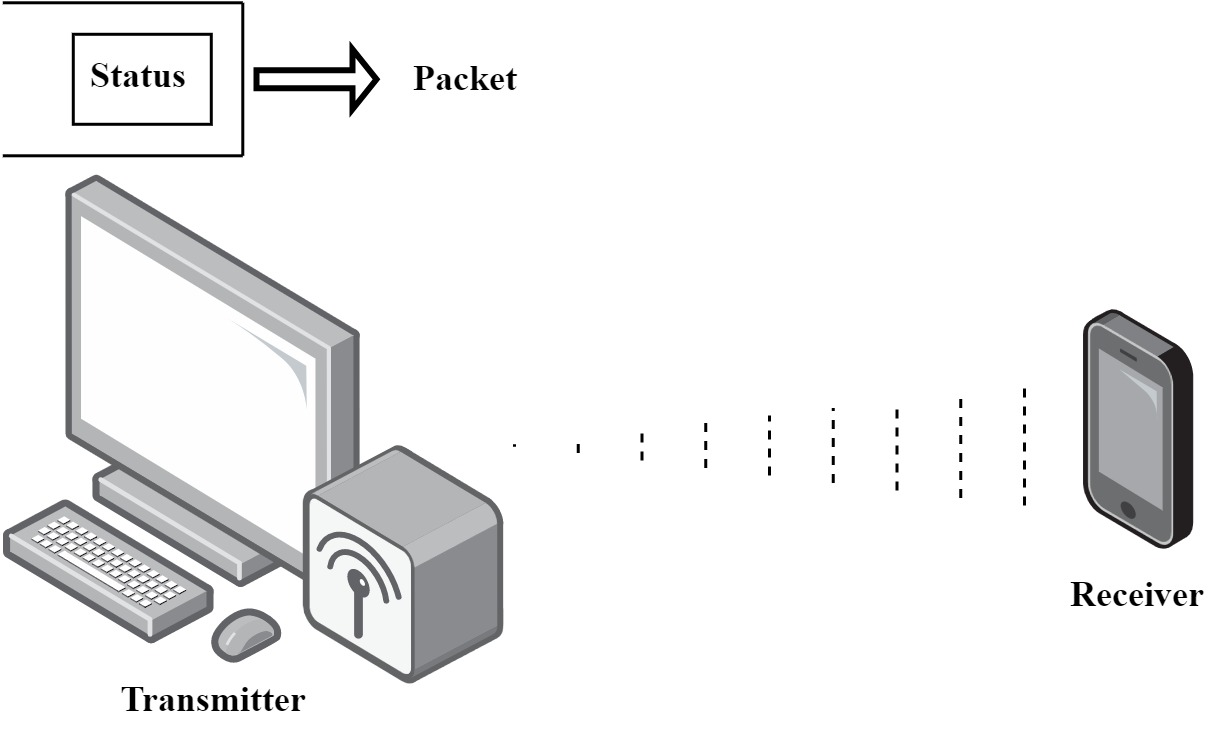}
\centering
\caption{Illustration of our considered system where the transmitter generates status updates, encodes the updates into packets, and transmits the packets to the receiver.}\label{fig:system_model}
\end{figure}

\subsection{System Description}

We consider a point-to-point wireless communication system, as depicted in Fig.~\ref{fig:system_model}, where a transmitter transmits packets to a receiver. In this system, the transmitter generates status updates according to a Poisson process with the rate of $\lambda$. Then the transmitter encodes the generated status updates into packets and transmits the encoded packets to the receiver. We assume that each status update contains a fixed number of bits, denoted by $L$. Once the transmitter generates a status update, it encodes these $L$-bit information into a packet with the blocklength of $M$ channel use (c.u.) and transmits this packet to the receiver. For this transmission, we define the coding rate, $R$, as the ratio between the number of bits in the status update and the blocklength of the transmitted packet, i.e., $R=L/M$.

To ensure the freshness of packets, we assume that the LCFS queuing policy is adopted and there is no feedback signal sent by the receiver. When a new status update is generated and the transmitter is in the idle state, it encodes the status update into a packet and transmits this packet to the receiver. In this system, we examine three packet management strategies, detailed as follows:
\begin{enumerate}
\item NPNB strategy: If a new status update is generated at the transmitter whilst the transmitter is transmitting a packet, the transmitter discards the new status update and keeps the current transmission. After transmitting the current packet, the transmitter waits for a new status update for transmission.
\item NPOB strategy: If a new status update is generated at the transmitter whilst the transmitter is transmitting a packet, the transmitter stores the newly generated status update in the buffer and keeps the current transmission. After transmitting the current packet, the transmitter immediately removes the status update from the buffer and transmits this status update to the receiver.
    %transmits the status update stored in the buffer and drops this status update from the buffer.
\item Preemption strategy: If a new status update is generated at the transmitter whilst the transmitter is transmitting a packet, the transmitter discards the current transmission and starts to transmit the newly generated status update. 
    %Preemption is considered such that the newly generated status update always preempts the transmission of the current status update. Specifically, if a new status update is generated at the transmitter and it is transmitting a status update, the transmitter discards the current transmission and starts to transmit the new status update.
\end{enumerate}

We assume an additive white Gaussian noise (AWGN) channel between the transmitter and the receiver, which is a standard assumption in the literature, e.g., \cite{WangGC2019,Cao2021AgeDelay}. According to \cite{Polyanskiy2010}, the block error rate for the AWGN channel using finite block length coding can be approximated as
\begin{align}\label{eq:blockerrrate}
\epsilon(l,m,\gamma) = Q\left(\frac{\frac{1}{2}\log_2(1+\gamma)-\frac{l}{m}}{\log_2(e)\sqrt{\frac{1}{2m}\left(1-\frac{1}{(1+\gamma^2)}\right)}}\right),
\end{align}
where $l$ is the number of bits in the status update, $m$ is the blocklength of the transmitted packet, $\gamma$ is the received signal-to-noise ratio (SNR) at the receiver, and $Q(x)=\int_{x}^{\infty}\frac{1}{\sqrt{2\pi}}e^{-\frac{t^2}{2}}dt$ is the $Q$-function. The block error rate expression in \eqref{eq:blockerrrate} is very tight for short packet communications when $m\geq100$ \cite{Polyanskiy2010}. For the packet transmission in our considered system, we define the transmission time for each c.u. as the unit time for analytical simplicity.
%For the packet transmission in our considered system, we denote $T_u$ as the transmission time for each c.u.. To facilitate our analysis, we define $T_u$ as the unit time.

\subsection{Formulation of AoI Penalty}

\begin{figure*}[t!]
\subfigure[]{
\begin{minipage}[b]{0.32\textwidth}
\centering
\includegraphics[width=1\columnwidth]{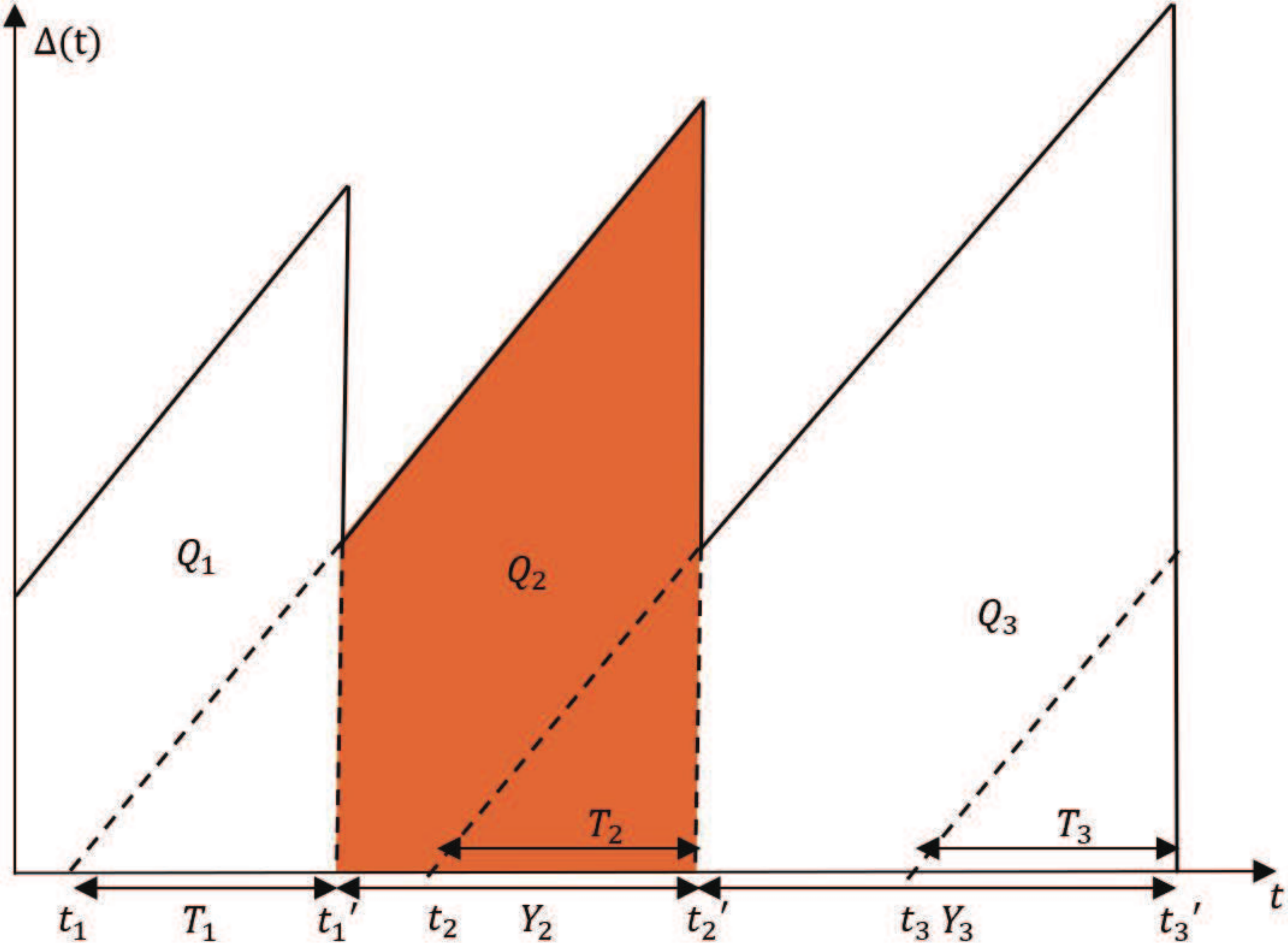}\label{fig:AoI1}
%\vspace{-1em}
\end{minipage}}
\subfigure[]{
\begin{minipage}[b]{0.32\textwidth}
\centering
\includegraphics[width=1\columnwidth]{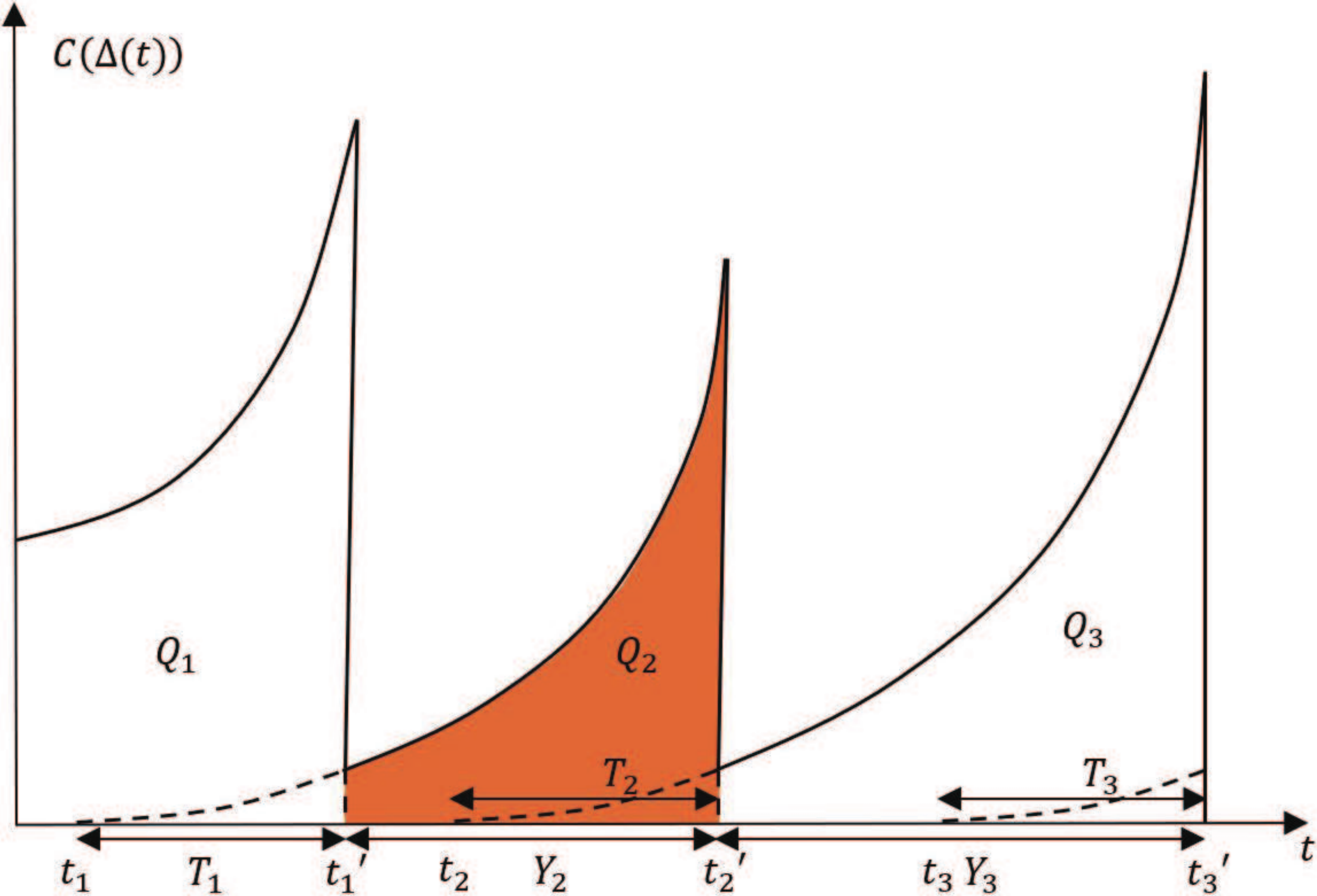}\label{fig:AoI2}
%\vspace{-1em}
\end{minipage}}
\subfigure[]{
\begin{minipage}[b]{0.32\textwidth}
\centering
\includegraphics[width=1\columnwidth]{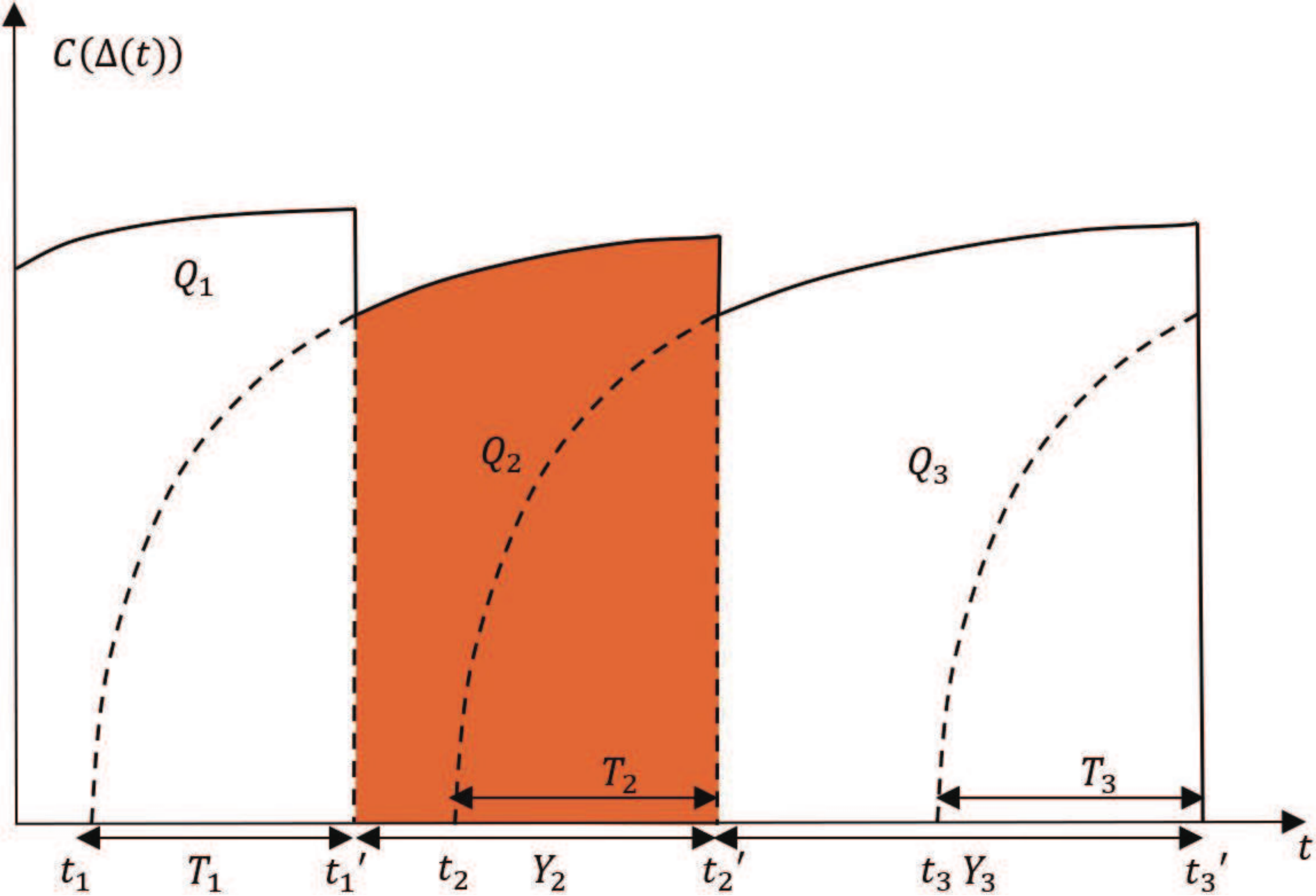}\label{fig:AoI3}
%\vspace{-1em}
\end{minipage}}
%\vspace{-0.5em}
\caption{Illustration of three AoI penalty functions with (a) linear AoI, (b) $\alpha>0$ and $\beta>0$, and (c) $\alpha<0$ and $\beta<0$.}\label{fig:AoI}
%\vspace{-1em}
\end{figure*}

In this subsection, we formulate the expression for the average AoI penalty of the considered system. To this end, we first denote $\Delta(t)$ as the AoI at time $t$. Fig.~\ref{fig:AoI1} plots a sample variation $\Delta(t)$ as a function of $t$. We assume that the observation begins at $t=0$ with the AoI of $\Delta(0)$. From Fig.~\ref{fig:AoI1}, we express the AoI at time $t$ as
\begin{align}\label{eq:2}
\Delta(t) = t - u(t),
\end{align}
where $u(t)$ is the generation time of the most recently received status update at the receiver at time $t$.

Apart from the linear AoI, we also adopt the $\alpha$-$\beta$ AoI penalty function \cite{Hu2022Penalty} in this work to characterize the level of the receiver's dissatisfaction on outdated data. Mathematically, this function is given by
\begin{align}
C\left(\Delta\left(t\right)\right)=\beta\left(e^{\alpha\Delta\left(t\right)}-1\right).
\end{align}
We find that %the $\alpha$-$\beta$ AoI penalty function,
$C\left(\Delta\left(t\right)\right)$ is an exponential-shape function when $\alpha>0$ and $\beta>0$, as shown in Fig.~\ref{fig:AoI2}, or a logarithmic-shape function when $\alpha<0$ and $\beta<0$, as shown in Fig.~\ref{fig:AoI3} \cite{Hu2022Penalty}.
%Moreover, 
We note that the $\alpha$-$\beta$ AoI penalty function is a generalized function as the linear AoI is a special case of $C\left(\Delta\left(t\right)\right)$ with $\lim\alpha\rightarrow0$ and $\beta=1/\alpha$.

We denote $P_{j}$ as the $j$th successfully received packet generated after time $t=0$, $j=1,2,\cdots$. We then denote $Y_j$ as the time interval from the received time of $P_{j-1}$ to the received time of $P_{j}$ and denote $T_j$ as the time interval from the generation time of $P_{j}$ to the received time of ${P}_{j}$. Thus, we obtain
\begin{align}
Y_j=t_{j}'-t_{j-1}'~~\textrm{and}~~T_j=t_j'-t_j,
\end{align}
%and
%\begin{align}    
%T_j=t_j'-t_j,
%\end{align}
where $t_j$ is the generation time of $P_{j}$ and $t_j'$ is the time that $P_{j}$ is received.

According to \cite{Hu2022Penalty}, the average AoI penalty is given by
\begin{align}\label{eq:newAoIevo}
\overline{C}=\frac{\E[Q_j]}{\E[Y_j]},
\end{align}
where $Q_j$ is the area under curve in Fig.~\ref{fig:AoI}. In \eqref{eq:newAoIevo}, the area $Q_j$ is calculated as
\begin{align}\label{eq:Qj}
Q_j &=\int_{t_{j-1}'}^{t_j'}C(\Delta(t))dt\notag\\
    &=\int_{0}^{Y_j+T_{j-1}}C(\Delta(t))dt-\int_{0}^{T_{j-1}}C(\Delta(t))dt\notag\\
    &=\frac{\beta}{\alpha}e^{\alpha T_{j-1}}\left(e^{\alpha Y_j}-1\right)-\beta Y_j.
\end{align}
By substituting \eqref{eq:Qj} into \eqref{eq:newAoIevo}, and using the fact that $Y_j$ is independent of $T_{j-1}$, we obtain
\begin{align}\label{eq:aveC}
\overline{C}
%&=\frac{\beta\E\left[e^{\alpha T_{j-1}}\left(e^{\alpha Y_j}-1\right)\right]}{\alpha\E\left[Y_j\right]}-\beta\notag\\
=\frac{\beta\E\left[e^{\alpha T_{j-1}}\right]\left(\E\left[e^{\alpha Y_j}\right]-1\right)}{\alpha\E\left[Y_j\right]}-\beta.
\end{align}
We next analyze $\overline{C}$ achieved by three strategies.

\section{Analysis of Average AoI Penalty}\label{sec:Derivation}

In this section, we derive the closed-form expressions for the average AoI penalty achieved by the three packet management strategies. We first focus on the NPNB strategy and present its average AoI penalty in the following theorem.
\begin{Theorem}\label{Theorem:1}
Considering the NPNB strategy with the received SNR $\gamma$, the closed-form expression for the average AoI penalty is derived as
\begin{align}\label{eq:expreAoIpar}
\overline{C}_{\mathrm{NPNB}}=\frac{\beta e^{\alpha M}\left(\lambda^2\left(1-\epsilon_{\gamma}\right)\left(e^{\alpha M}-1\right)+\alpha\right)}{\alpha\left(\lambda M+1\right)\left(\lambda\left(1-\epsilon_{\gamma}e^{\alpha M}\right)-\alpha\right)}-\beta,
\end{align}
where $\epsilon_{\gamma}=\epsilon(L,M,\gamma)$ and $\epsilon(\cdot,\cdot,\cdot)$ is given in \eqref{eq:blockerrrate}.
\begin{IEEEproof}
Based on \eqref{eq:aveC}, we need to derive $\E[e^{\alpha T_{j-1}}]$, $\E\left[e^{\alpha Y_j}\right]$, and $\E[Y_j]$ to obtain the average AoI penalty. Since $T_{j-1} = M$ for the NPNB strategy, we first obtain
\begin{align}\label{eq:T1}
\E\left[e^{\alpha T_{j-1}}\right] = e^{\alpha M}.
\end{align}

We then derive $\E\left[e^{\alpha Y_j}\right]$ and $\E[Y_j]$. Here, we denote $H_j$ as the number of packets transmitted to the receiver after $P_{j-1}$ is received but by $P_j$ is received, and $P_{j,h}$ as the $h$th packet transmitted to the receiver after $P_{j-1}$ is received, where $P_{j,H_j}=P_j$. We note that
\begin{align}\label{eq:YjS2e}
Y_j=\sum_{h=1}^{H_j}B_{j,h},
\end{align}
where $B_{j,h}$ is the time interval from the time that $P_{j,h-1}$ is transmitted to the receiver to the time that $P_{j,h}$ is transmitted to the receiver. Here, we note that if $h = 1$, $P_{j,0}=P_{j-1}$. We find that $H_j$ follows a geometry distribution, where the probability mass function (PMF) of $H_j$ is given by
\begin{align}\label{eq:Kdistricase1}
\mathrm{Pr}\left[H_j=H\right]=\left(1-\epsilon_{\gamma}\right)\epsilon_{\gamma}^{H-1}.
\end{align}
In \eqref{eq:YjS2e}, $B_{j,h}$ is calculated by
\begin{align}\label{eq:Bjkcase1}
B_{j,h}=V_{j,h} + M,
\end{align}
where $V_{j,h}$ is the time that the transmitter is in the idle state during $B_{j,h}$. Since $V_{j,k}$ follows an exponential distribution, we obtain
\begin{align}\label{eq:Vjkcase1}
\E\left[e^{\alpha B_{j,h}}\right]&=e^{\alpha M}\E\left[e^{\alpha V_{j,h}}\right]
%=e^{\alpha M}\int_0^{\infty}\lambda e^{-\lambda t}e^{\alpha t}dt\notag\\
=e^{\alpha M}\frac{\lambda}{\lambda-\alpha}.
\end{align}
By substituting \eqref{eq:Vjkcase1} and \eqref{eq:Kdistricase1} into \eqref{eq:YjS2e}, we obtain $\E\left[e^{\alpha Y_j}\right]$ as
\begin{align}\label{eq:eaYj1}
\E\left[e^{\alpha Y_j}\right]&= \sum_{h=1}^{\infty}\left(1-\epsilon_{\gamma}\right)\epsilon_{\gamma}^{h-1}\E\left[e^{\alpha B_{j,h}}\right]^{h}\notag\\
&=\frac{\lambda\left(1-\epsilon_{\gamma}\right)e^{\alpha M}}{\lambda(1-\epsilon_{\gamma}e^{\alpha M})-\alpha}.
\end{align}

Finally, based on \eqref{eq:YjS2e} and \eqref{eq:Bjkcase1}, we calculate $\E[Y_j]$ as
\begin{align}\label{eq:Yj1}
\E\left[Y_j\right]=\frac{\lambda M+1}{\lambda\left(1-\epsilon_{\gamma}\right)}.
\end{align}
By substituting \eqref{eq:T1}, \eqref{eq:eaYj1}, and \eqref{eq:Yj1} into \eqref{eq:aveC}, we obtain the final result in \eqref{eq:expreAoIpar}, which completes the proof.
\end{IEEEproof}
\end{Theorem}

We find from Theorem~\ref{Theorem:1} that in order to guarantee the existence of a finite average AoI penalty, $\alpha$ needs to satisfy two conditions, namely, $\alpha<-\frac{\log\epsilon_{\gamma}}{M}$ and $\lambda\left(1-\epsilon_{\gamma}e^{\alpha M}\right)>\alpha$. Otherwise, the average AoI penalty goes to infinity in the NPNB strategy.

We then derive and present the closed-form expression for the average AoI penalty achieved by the NPOB strategy in the following theorem.
\begin{Theorem}\label{Theorem:2}
Considering the NPOB strategy with the received SNR $\gamma$, the closed-form expression for the average AoI penalty is derived as
\begin{align}\label{eq:expreAoIpar1}
&\overline{C}_{\mathrm{NPB}}=\frac{\beta\lambda e^{\alpha M}\left(1-\epsilon_{\gamma}\right)}{\alpha\left(\lambda M\!+\!e^{-\lambda M}\right)}\!\left(e^{-\lambda M}+\frac{\lambda}{\lambda-\alpha}\left(1-e^{(\alpha-\lambda) M}\right)\!\right)\notag\\
&\times\!\left(\!\frac{\left(1-\epsilon_{\gamma}\right)e^{\alpha M}\left(\lambda-\alpha(1-e^{-\lambda M})\right)}{\lambda\left(1-e^{\alpha M}\epsilon_{\gamma}\right)\!-\!\alpha\left(1-\epsilon_{\gamma}(e^{\alpha M}-e^{(\alpha-\lambda) M})\right)}-1\!\right)\!-\!\beta.
\end{align}
\begin{IEEEproof}
Again, we need to derive $\E\left[e^{\alpha T_{j-1}}\right]$, $\E\left[e^{\alpha Y_j}\right]$, and $\E[Y_j]$ to obtain the average AoI penalty. We first derive $\E\left[e^{\alpha T_{j-1}}\right]$ as
\begin{align}\label{eq:Tjcase1}
\E\left[e^{\alpha T_{j-1}}\right]=\E\left[e^{\alpha\left(M+W_{j-1}\right)}\right]
=e^{\alpha M}\E\left[e^{\alpha W_{j-1}}\right],
\end{align}   
where $W_{j-1}$ is the waiting time of $P_{j-1}$ in the buffer. We note that $W_{j-1}=0$ when $P_{j-1}$ is generated after the transmission of $P_{j-2}$ and it follows an exponential distribution when $P_{j-1}$ is generated during the transmission of $P_{j-2}$. We then calculate $\E\left[e^{\alpha W_{j-1}}\right]$ as
\begin{align}\label{eq:Wjcase1}
\E\left[e^{\alpha W_{j-1}}\right]&=\int_{0}^{M}e^{\alpha t}\lambda e^{-\lambda t}dt+\int_{M}^{\infty}\lambda e^{-\lambda t}dt\notag\\
&=e^{-\lambda M}+\frac{\lambda}{\lambda-\alpha}\left(1-e^{\left(\alpha-\lambda\right)M}\right).
\end{align}
By substituting \eqref{eq:Wjcase1} into \eqref{eq:Tjcase1}, we obtain $\E[e^{\alpha T_{j-1}}]$.

We then derive $\E\left[e^{\alpha Y_j}\right]$ and $\E[Y_j]$. We note that the probability of $V_{j,h} = 0 $ is $\mathrm{Pr}\left[V_{j,h} = 0\right]= 1-e^{-\lambda M}$ for the NPOB strategy. Hence, we calculate $\E\left[e^{\alpha V_{j,h}}\right]$ and $\E[V_{j,h}]$ as
\begin{align}\label{eq:eVjhcase2}
\E\left[e^{\alpha V_{j,h}}\right]&=1-e^{-\lambda M}+e^{-\lambda M} \int_0^{\infty}e^{\alpha t}\lambda e^{-\lambda t}dt\notag\\
&=1+\frac{\alpha e^{-\lambda M}}{\lambda-\alpha}.
\end{align}
and 
\begin{align}\label{eq:Vjhcase2}
\E\left[V_{j,h}\right]=e^{-\lambda M}\int_0^{\infty}t\lambda e^{-\lambda t}dt = \frac{e^{-\lambda M}}{\lambda},
\end{align}
respectively. By substituting \eqref{eq:eVjhcase2} and \eqref{eq:Vjhcase2} into \eqref{eq:Bjkcase1}, and combining the result with \eqref{eq:YjS2e}, we obtain $\E\left[e^{\alpha Y_j}\right]$ as
\begin{align}\label{eq:EeYjcase2}
\E\left[e^{\alpha Y_j}\right]=\frac{(1-\epsilon_{\gamma})e^{\alpha M}\left(\lambda-\alpha(1-e^{-\lambda M})\right)}{\lambda\left(1-e^{\alpha M}\epsilon_{\gamma}\right)\!-\!\alpha\left(1-\epsilon_{\gamma}(e^{\alpha M}-e^{\left(\alpha-\lambda\right)M})\right)}
\end{align}
and obtain $\E\left[Y_j\right]$ as
\begin{align}\label{eq:EYjcase2}
\E\left[Y_j\right]=\frac{\lambda M+e^{-\lambda M}}{\lambda(1-\epsilon_{\gamma})}.
\end{align}
By substituting \eqref{eq:Tjcase1}, \eqref{eq:EeYjcase2}, and \eqref{eq:EYjcase2} into \eqref{eq:aveC}, we obtain the final result in \eqref{eq:expreAoIpar1}, completing this proof.
\end{IEEEproof}
\end{Theorem}

We find from Theorem~\ref{Theorem:2} that in order to guarantee the existence of a finite average AoI penalty, $\alpha$ needs to satisfy two conditions, namely, $\alpha<-\frac{\log\epsilon_{\gamma}}{M}$ and $(\lambda-\alpha)\left(1-\epsilon_{\gamma}e^{\alpha M}\right)>\alpha\epsilon_{\gamma}e^{(\alpha-\lambda) M}$. Otherwise, the average AoI penalty goes to infinity in the NPOB strategy.

We note that the information freshness is usually improved by eliminating the waiting time of the transmitter for a new packet and the waiting time of a packet in the buffer, which is the zero-waiting policy. Under this policy, i.e., $\lambda\rightarrow\infty$, we find that the average AoI penalty achieved by the NPNB strategy is same as the average AoI penalty achieved by the NPOB strategy, which is given by
\begin{align}
\overline{C}_{\mathrm{ZW}}=\frac{\beta e^{\alpha M}\left(1-\epsilon_{\gamma}\right)\left(e^{\alpha M}-1\right)}
{\alpha M\left(1-\epsilon_{\gamma}e^{\alpha M}\right)}-\beta,
\end{align}
where $\alpha<-\frac{\log\epsilon_{\gamma}}{M}$ needs to be satisfied to guarantee the existence of a finite average AoI penalty.

We further derive the average AoI penalty achieved by the preemption strategy in the following Theorem.
\begin{Theorem}\label{Theorem:3}
Considering the preemption strategy with the received SNR $\gamma$, the closed-form expression for the average AoI penalty is derived as
\begin{align}\label{eq:expreAoIpar2}
\overline{C}_{\mathrm{P}}=\frac{\beta\lambda e^{\left(\alpha-\lambda\right) M}\left(1\!-\!\epsilon_{\gamma}\right)\left(\alpha e^{\lambda M}-\lambda \epsilon_{\gamma}\left(e^{\alpha M}-1\right)\right)}{\alpha\left(\left(\lambda e^{\alpha M}-\alpha e^{\lambda M}\right)-\lambda\epsilon_{\gamma}\right)}-\beta.
\end{align}
\begin{IEEEproof}
To obtain the average AoI penalty, $\E\left[e^{\alpha T_{j-1}}\right]$, $\E\left[e^{\alpha Y_j}\right]$, and $\E\left[Y_j\right]$ need to be derived. Since $T_{j-1} = M$ for the preemption strategy, we obtain $\E\left[e^{\alpha T_{j-1}}\right]$ as
\begin{align}\label{eq:T2}
\E\left[e^{\alpha T_{j-1}}\right]=e^{\alpha M}.
\end{align}

We then derive $\E\left[e^{\alpha Y_j}\right]$ and $\E\left[Y_j\right]$. For the preemption strategy, we note that
\begin{align}\label{eq:Bjkcase2}
B_{j,h}=V_{j,h}+\sum_{k=0}^{K_{j,h}}S_{j,h,k}+M,
\end{align}
where $K_{j,h}$ is the preemption times of the status update of $P_{j,h}$ and $S_{j,h,k}$ is the transmission time of the $k$th preempted status update. We note that $K_{j,h}$ follows a geometry distribution, i.e., $\mathrm{Pr}(K_{j,h}=K) = (1-e^{-\lambda M})^{K}e^{-\lambda M}$. Hence, we calculate $\E\left[e^{\alpha B_{j,h}}\right]$ and $\E\left[B_{j,h}\right]$ as
\begin{align}\label{eq:EeBjhcase3}
\E\left[e^{\alpha B_{j,h}}\right]&=\frac{\lambda e^{\alpha M}}{\lambda-\alpha}\sum_{k= 0}^{\infty}e^{-\lambda M}\left(\frac{\lambda\left(e^{\lambda M}-e^{\alpha M}\right)}{\left(\lambda-\alpha\right)e^{\lambda M}}\right)^k\notag\\
&=\frac{\lambda e^{\alpha M}}{\lambda e^{\alpha M}-\alpha e^{\lambda M}},
\end{align}
and
\begin{align}\label{eq:EBjhcase3}
\E\left[B_{j,h}\right]=\frac{1}{\lambda}+M+\E\left[K_{j,h}\right]
\E\left[S_{j,h,k}\right]=\frac{1}{\lambda e^{-\lambda M}},
\end{align}
respectively. By substituting \eqref{eq:EeBjhcase3} and \eqref{eq:EBjhcase3} into \eqref{eq:YjS2e}, we obtain $\E\left[e^{\alpha Y_j}\right]$ and $\E\left[Y_j\right]$ as
\begin{align}\label{eq:EeYjcase3}
\E\left[e^{\alpha Y_j}\right]=\frac{\lambda e^{\alpha M}(1-\epsilon_{\gamma})}{\lambda e^{\alpha M}-\alpha e^{\lambda M}-\lambda \epsilon_{\gamma}},
\end{align}
and
\begin{align}\label{eq:EYjcase3}
\E\left[Y_j\right]=\frac{1}{\lambda e^{-\lambda M}\left(1-\epsilon_{\gamma}\right)},
\end{align}
respectively. By substituting \eqref{eq:T2}, \eqref{eq:EeYjcase3}, and \eqref{eq:EYjcase3} into \eqref{eq:aveC}, we obtain the final result in \eqref{eq:expreAoIpar2}. This completes the proof.
\end{IEEEproof}
\end{Theorem}

We find from Theorem~\ref{Theorem:3} that in order to guarantee the existence of a finite average AoI penalty, $\alpha$ needs to satisfy two conditions, namely, $\alpha<\lambda$ and $\left(\lambda e^{\alpha M}-\alpha e^{\lambda M}\right)>\lambda\epsilon_{\gamma}$. Otherwise, the average AoI penalty goes to infinity in the preemption strategy.

So far, we have obtained the expressions for the average AoI penalty achieved by the NPNB strategy, the NPOB strategy, and the preemption strategy in Theorems \ref{Theorem:1}, \ref{Theorem:2}, and \ref{Theorem:3}, respectively. Since the block error rate is the function of the packet blocklength, there would exists the optimal coding rate to minimize the average AoI penalty for three strategies. Due to the complex form of the block error rate given in \eqref{eq:blockerrrate}, it is difficult to derive closed-form expressions for the optimal packet blocklength for the three strategies. Thus, we resort to numerical methods, e.g., numerical search methods, to find the optimal packet blocklengths, which are the solutions to the equation $\frac{d\overline{C}}{dM}=0$ for the three strategies.

\section{Numerical Results}\label{sec:Numerical}

In this section, we present numerical results to evaluate the impact of various parameters, including the coding rate, status update generation rate, and the value of $\alpha$ on the average AoI penalty, achieved by three strategies in our considered system. We then examine how the value of $\alpha$ affects the optimal coding rate in the system.

\begin{figure}[!t]
\centering
\includegraphics[width=0.85\columnwidth]{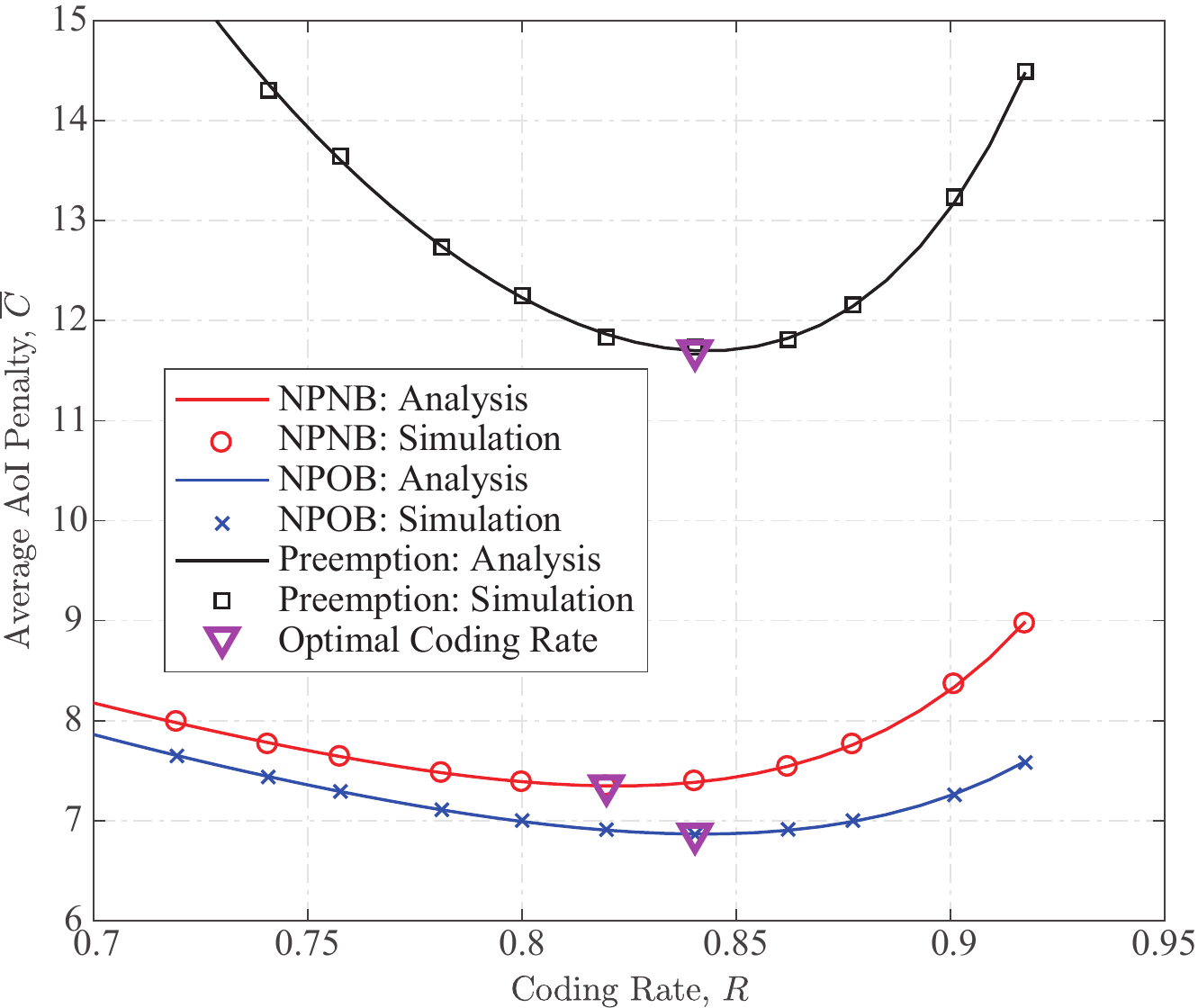}
%\vspace{-0.5em}
\caption{Average AoI penalty versus the coding rate, $R$, with $\lambda = 0.01$, $\gamma=3$, and $L=100$ for $C\left(\Delta(t)\right)=10\left(e^{0.002\Delta(t)}-1\right)$.}\label{fig:CodeR}
\vspace{-1.5em}
\end{figure}

Fig.~\ref{fig:CodeR} plots the average AoI penalty versus the coding rate, $R$. We first observe that the analytical average AoI penalty precisely matches the simulation results, which demonstrates the correctness of our analytical results in Theorems \ref{Theorem:1}, \ref{Theorem:2}, and \ref{Theorem:3}. We then observe that for three strategies, the average AoI penalty first decreases and then increases when $R$ increases. % This implies that there is an optimal value of $\lambda_h$ to minimize the average AoI of the MEC system.
This observation is due to the fact that the increase in $R$ has a two-fold effect on the average AoI penalty through the packet blocklength and the block error rate. When $R$ is small, the increase in $R$ leads to the smaller packet blocklength, which decreases the average AoI penalty. When $R$ is large and exceeds a certain threshold, the increase in $R$ brings about a significant increase in the block error rate, thereby increasing the average AoI penalty.

\begin{figure}[t]
\subfigure[]{
\begin{minipage}[b]{0.225\textwidth}
\centering
\includegraphics[width=1\columnwidth]{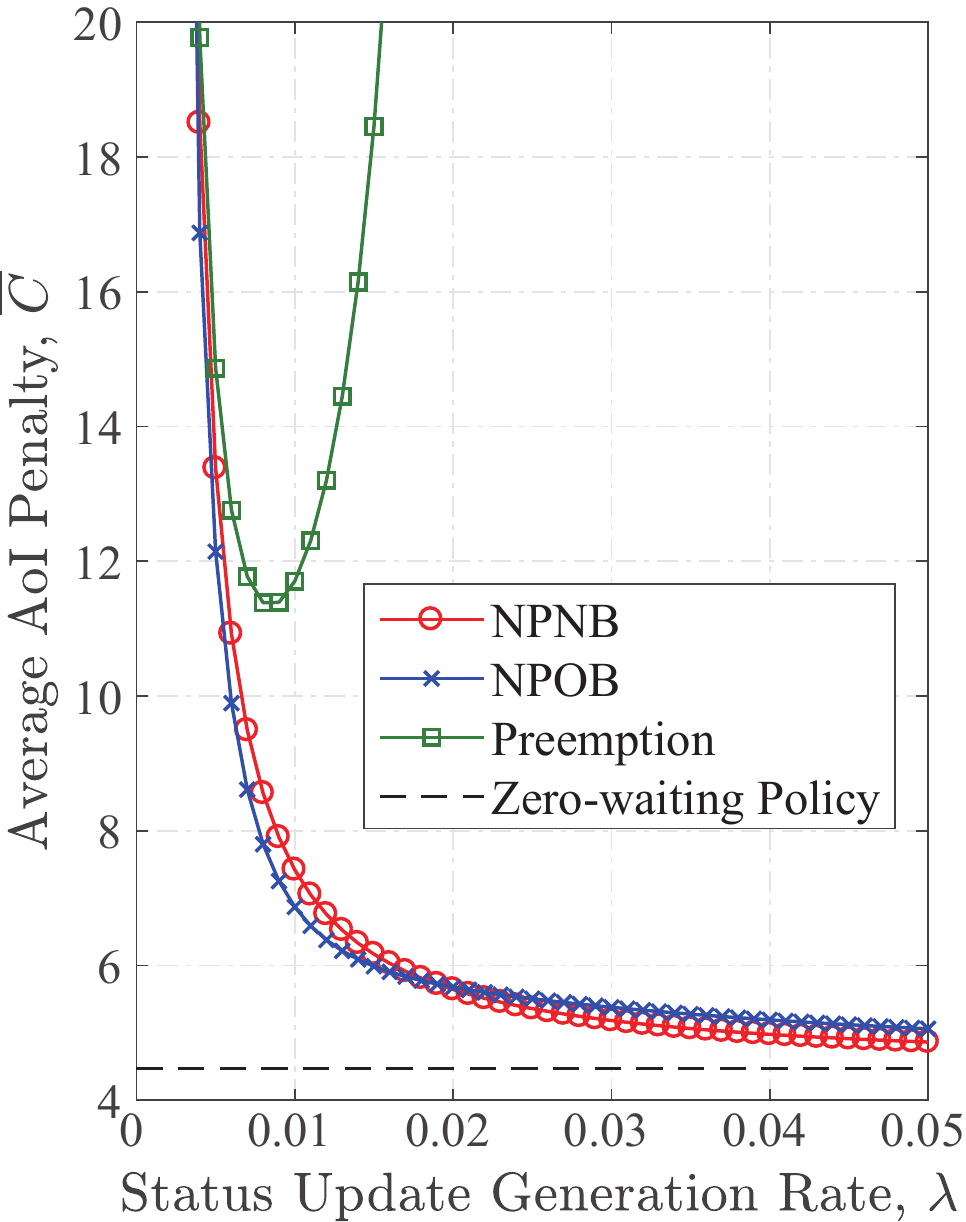}\label{fig:l1}
%\vspace{-1em}
\end{minipage}}
\subfigure[]{
\begin{minipage}[b]{0.225\textwidth}
\centering
\includegraphics[width=1\columnwidth]{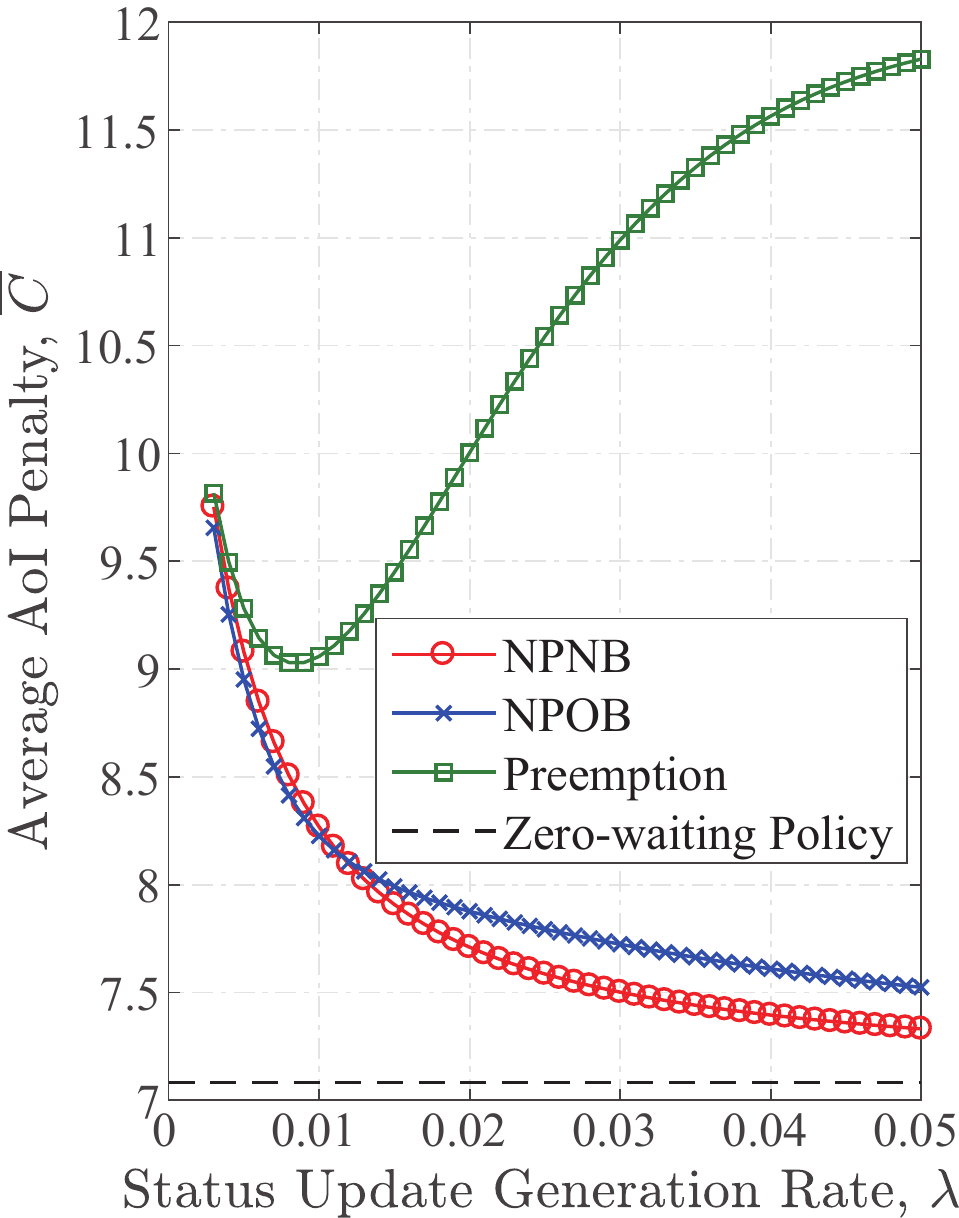}\label{fig:l2}
%\vspace{-1em}
\end{minipage}}
\vspace{-0.5em}
\caption{Average AoI penalty versus the status update generation rate, $\lambda$, with $R = 0.85$, $\gamma=3$, and $L=100$, for (a) $C\left(\Delta(t)\right)=10\left(e^{0.002\Delta(t)}-1\right)$ and (b) $C\left(\Delta(t)\right)=-12\left(e^{-0.005\Delta(t)}-1\right)$.}\label{fig:l}
\vspace{-1.5em}
\end{figure}

Fig.~\ref{fig:l} plots the average AoI penalty versus the status update generation rate, $\lambda$. We first observe that for both the NPNB and NPOB strategies, the average AoI penalty decreases monotonically when $\lambda$ increases, which approaches the average AoI penalty under the zero-waiting policy when $\lambda$ is large. This is because that the increase in $\lambda$ decreases the transmitter's waiting time for a new status update, which in turn decreases the average AoI penalty. We then observe that the NPOB strategy achieves a lower average AoI penalty than the NPNB strategy when $\lambda$ is small, but a larger average AoI penalty when $\lambda$ is large. This is because that when $\lambda$ is small, the NPOB strategy has a higher update frequency, which deceases the average AoI penalty. When $\lambda$ is large and exceeds a threshold, the priority of this higher update frequency reduces and the transmitted packets are outdated comparing to the NPNB strategy, leading to the higher average AoI penalty. We further observe that the average AoI first decreases and then increases when $\lambda$ increases in the preemption strategy. This observation is due to the fact that the increase in $\lambda$ has a two-fold effect on the average AoI penalty. When $\lambda$ is small, the increase in $\lambda$ reduces the transmitter's waiting time after a successful transmission, which decreases the average AoI penalty. When $\lambda$ exceeds a certain threshold, the increase in $\lambda$ leads to a significant increase in the probability that a status update is preempted, thereby increasing the average AoI penalty. In addition, we find that non-preemption strategies, including the NPNB and NPOB strategies, achieve a better AoI performance than the preemption strategy.

\begin{figure}[!t]
\centering
\includegraphics[width=0.85\columnwidth]{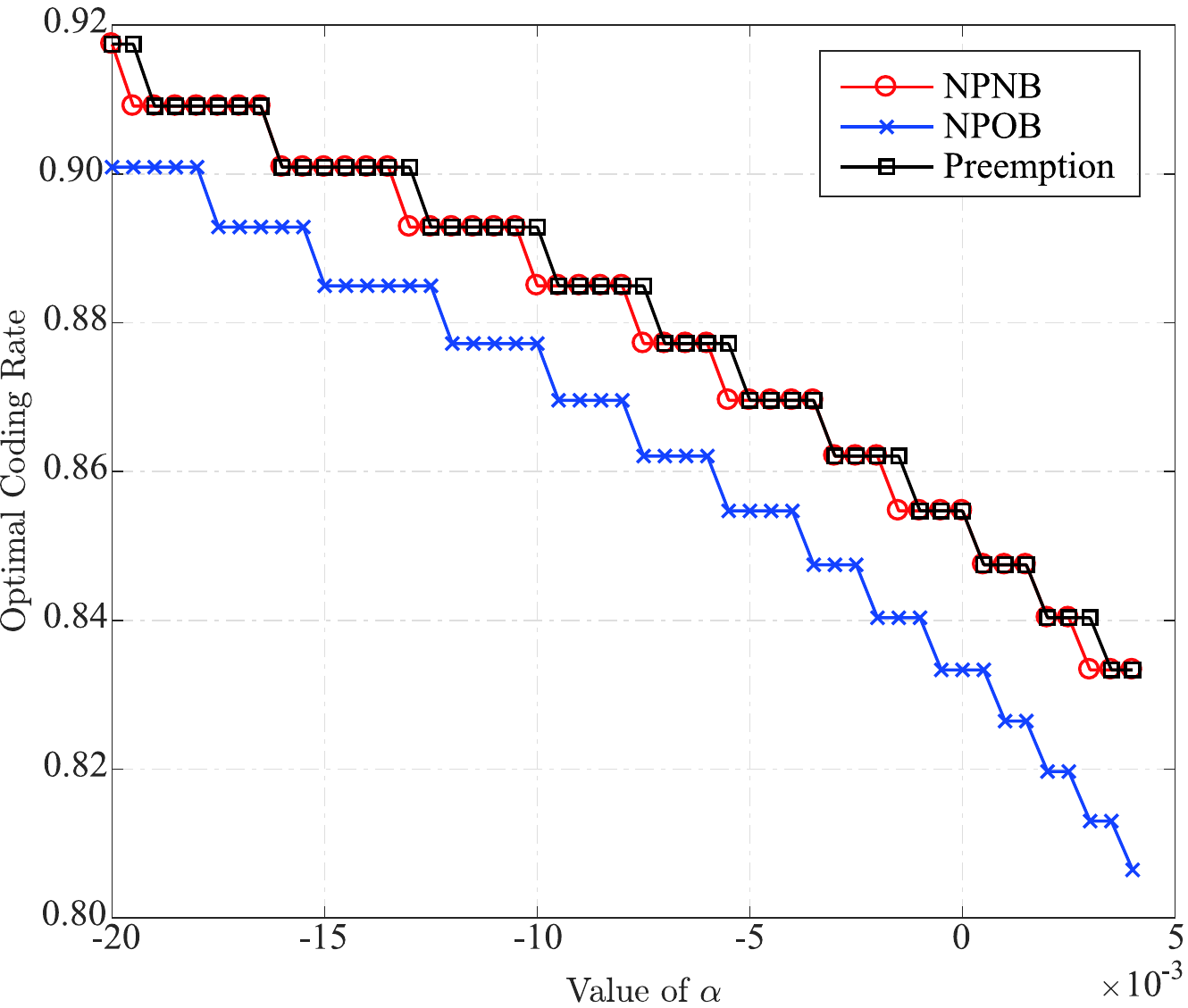}
%\vspace{-0.5em}
\caption{The optimal coding rate of the system versus the value of $\alpha$, with $\lambda = 0.01$, $\gamma=3$, $L=100$, and $\beta=\frac{1}{\alpha}$.}\label{fig:4ps}
\vspace{-1.5em}
\end{figure}

Fig.~\ref{fig:4ps} plots the optimal coding rate versus the value of $\alpha$. We observe that the optimal coding rate decreases as the value of $\alpha$ increases. Here, we note that a lower coding rate represents fewer transmission errors and higher transmission reliability. It implies that the system requires high reliability when having a relatively low tolerance to outdated data, but low reliability when having a relatively high tolerance to outdated data. It follows that the non-linear AoI penalty reflects not only the receiver’s dissatisfaction on outdated data but also the reliability requirement of short packet communications. Based on this observation, the value of $\alpha$ can be designed according to the transmission reliability requirement. 

\section{Conclusion}\label{sec:Conclusion}

This paper considered a point-to-point system where a transmitter generates packets and transmits them to a receiver via an unreliable channel. By considering non-linear $\alpha$-$\beta$ AoI penalty, we analyzed the average AoI penalty achieved by three packet management strategies, i.e., the NPNB strategy, the NPOB strategy, and the preemption strategy, based on a finite packet blocklength model. With simulation results, we demonstrated the accuracy of our analysis and showed that the NPNB strategy and NPOB strategies are more suitable than the preemption strategy for short packet communication systems by achieving a lower average AoI penalty. Moreover, we found that the system requires high transmission reliability when the receiver has a low tolerance to outdated data, but low transmission reliability when the receiver has a high tolerance.

\section*{Acknowledgment}

This work was funded by the Australian Research Council Discovery Project DP180104062.

\bibliographystyle{IEEEtran} %references to be listed in the order of appeareance
\bibliography{bibli}

\begin{thebibliography}{10}
\providecommand{\url}[1]{#1}
\csname url@samestyle\endcsname
\providecommand{\newblock}{\relax}
\providecommand{\bibinfo}[2]{#2}
\providecommand{\BIBentrySTDinterwordspacing}{\spaceskip=0pt\relax}
\providecommand{\BIBentryALTinterwordstretchfactor}{4}
\providecommand{\BIBentryALTinterwordspacing}{\spaceskip=\fontdimen2\font plus
\BIBentryALTinterwordstretchfactor\fontdimen3\font minus
  \fontdimen4\font\relax}
\providecommand{\BIBforeignlanguage}[2]{{%
\expandafter\ifx\csname l@#1\endcsname\relax
\typeout{** WARNING: IEEEtran.bst: No hyphenation pattern has been}%
\typeout{** loaded for the language `#1'. Using the pattern for}%
\typeout{** the default language instead.}%
\else
\language=\csname l@#1\endcsname
\fi
#2}}
\providecommand{\BIBdecl}{\relax}
\BIBdecl

\bibitem{Li2019}
C.~{Li}, N.~{Yang}, and S.~{Yan}, ``Optimal transmission of short-packet
  communications in multiple-input single-output systems,'' \emph{IEEE Trans.
  Veh. Technol.}, vol.~68, no.~7, pp. 7199--7203, Jul. 2019.

\bibitem{Sun2018}
X.~{Sun}, S.~{Yan}, N.~{Yang}, Z.~{Ding}, C.~{Shen}, and Z.~{Zhong}, ``Downlink
  {NOMA} transmission for low-latency short-packet communications,'' in
  \emph{Proc. IEEE Int. Commun. Conf.}, Kansas City, MO, May 2018, pp. 1--6.

\bibitem{Huang2019}
X.~{Huang} and N.~{Yang}, ``On the block error performance of short-packet
  non-orthogonal multiple access systems,'' in \emph{Proc. IEEE Int. Commun.
  Conf.}, Shanghai, China, May 2019, pp. 1--7.

\bibitem{Li2021}
C.~Li, C.~She, N.~Yang, and T.~Q.~S. Quek, ``Secure transmission rate of short
  packets with queueing delay requirement,'' \emph{IEEE Trans. Wireless
  Commun.}, pp. 203--218, Jan. 2022.

\bibitem{Kaul2011}
S.~{Kaul}, M.~{Gruteser}, V.~{Rai}, and J.~{Kenney}, ``Minimizing age of
  information in vehicular networks,'' in \emph{Proc. IEEE Conf. Sensor Ad Hoc
  Commun. Netw.}, Salt Lake City, UT, Jun. 2011, pp. 350--358.

\bibitem{Tang2020}
Z.~{Tang}, Z.~{Sun}, N.~{Yang}, and X.~{Zhou}, ``Age of information of
  multi-source systems with packet management,'' in \emph{Proc. IEEE Int.
  Commun. Conf.}, Dublin, Ireland, Jun. 2020, pp. 1--6.

\bibitem{Wang2020IoT}
Q.~{Wang}, H.~{Chen}, C.~{Zhao}, Y.~{Li}, P.~{Popovski}, and B.~{Vucetic},
  ``Optimizing information freshness via multiuser scheduling with adaptive
  {NOMA}/{OMA},'' \emph{IEEE Intern. of Things J.}, vol.~7, no.~9, pp.
  8178--8191, Sep. 2020.

\bibitem{Tang2021GC}
Z.~{Tang}, Z.~{Sun}, N.~{Yang}, and X.~{Zhou}, ``Age of information analysis of
  multi-user mobile edge computing systems,'' in \emph{Proc. IEEE Global
  Commun. Conf.}, Dec. 2021, pp. 1--6.

\bibitem{Wang2019a}
Q.~{Wang}, H.~{Chen}, Y.~{Li}, Z.~{Pang}, and B.~{Vucetic}, ``Minimizing age of
  information for real-time monitoring in resource-constrained industrial {IoT}
  networks,'' in \emph{Proc. IEEE Intern. Conf. Industr. Inform.}, Helsinki,
  Finland, Jul. 2019, pp. 1766--1771.

\bibitem{Kadota2016}
I.~{Kadota}, E.~{Uysal-Biyikoglu}, R.~{Singh}, and E.~{Modiano}, ``Minimizing
  the age of information in broadcast wireless networks,'' in \emph{Proc.
  Allerton Conf. on Commun. Control, and Comput.}, Monticello, IL, Sep. 2016,
  pp. 844--851.

\bibitem{Hsu2018}
Y.~{Hsu}, ``Age of information: Whittle index for scheduling stochastic
  arrivals,'' in \emph{Proc. IEEE Int. Sympos. Inf. Theory}, Vail, CO, Jun.
  2018, pp. 2634--2638.

\bibitem{Sun2019A}
J.~{Sun}, Z.~{Jiang}, S.~{Zhou}, and Z.~{Niu}, ``Optimizing information
  freshness in broadcast network with unreliable links and random arrivals: An
  approximate index policy,'' in \emph{Proc. IEEE Int. Conf. Comput. Commun.},
  Paris, France, May 2019, pp. 115--120.

\bibitem{Kosta2020}
A.~{Kosta}, N.~{Pappas}, A.~{Ephremides}, and V.~{Angelakis}, ``The cost of
  delay in status updates and their value: Non-linear ageing,'' \emph{IEEE
  Trans. Commun.}, vol.~68, no.~8, pp. 4905--4918, Apr. 2020.

\bibitem{Sun2016Fresh}
Y.~{Sun}, E.~{Uysal-Biyikoglu}, R.~{Yates}, C.~E. {Koksal}, and N.~B. {Shroff},
  ``Update or wait: How to keep your data fresh,'' in \emph{Proc. IEEE Int.
  Conf. Comput. Commun.}, Apr. 2016, pp. 1--9.

\bibitem{Zheng2019NL}
X.~{Zheng}, S.~{Zhou}, Z.~{Jiang}, and Z.~{Niu}, ``Closed-form analysis of
  non-linear age of information in status updates with an energy harvesting
  transmitter,'' \emph{IEEE Trans. Wireless Commun.}, vol.~18, no.~8, pp.
  4129--4142, Aug. 2019.

\bibitem{Klugel2019}
M.~{Kl$\mathrm{\ddot{u}}$gel}, M.~H. {Mamduhi}, S.~{Hirche}, and W.~{Kellerer},
  ``{AoI}-penalty minimization for networked control systems with packet
  loss,'' in \emph{Proc. IEEE Int. Conf. Comput. Commun.}, Paris, France, May
  2019, pp. 189--196.

\bibitem{tang2021whittle}
Z.~{Tang}, Z.~{Sun}, N.~{Yang}, and X.~{Zhou}, ``Whittle index based scheduling
  policy for minimizing the cost of age of information,'' \emph{IEEE Commun.
  Lett.}, vol.~26, no.~1, pp. 54--58, Jan. 2022.

\bibitem{Hu2022Penalty}
H.~{Hu}, K.~{Xiong}, Y.~{Lu}, B.~{Gao}, P.~{Fan}, and K.~B. {Letaief},
  ``$\alpha$-$\beta$ {AoI} penalty in wireless-powered status update
  networks,'' \emph{IEEE Internet Things J.}, vol.~9, no.~1, pp. 474--484,
  2022.

\bibitem{devassy2018delay}
R.~Devassy, G.~Durisi, G.~C. Ferrante, O.~Simeone, and E.~Uysal-Biyikoglu,
  ``Delay and peak-age violation probability in short-packet transmissions,''
  in \emph{Proc. IEEE Int. Sympos. Inf. Theory}, Vail, CO, Jun. 2018, pp.
  2471--2475.

\bibitem{Devassy2019}
R.~{Devassy}, G.~{Durisi}, G.~C. {Ferrante}, O.~{Simeone}, and E.~{Uysal},
  ``Reliable transmission of short packets through queues and noisy channels
  under latency and peak-age violation guarantees,'' \emph{IEEE J. Select.
  Areas Commun.}, vol.~37, no.~4, pp. 721--734, Apr. 2019.

\bibitem{Basnay2021}
C.~M.~W. Basnayaka, D.~N.~K. Jayakody, T.~D.~P. Perera, and M.~V. Ribeiro,
  ``Age of information in an {URLLC}-enabled decode-and-forward wireless
  communication system,'' in \emph{Proc. IEEE Veh. Techn. Conf.}, Helsinki,
  Finland, Apr. 2021, pp. 1--6.

\bibitem{WangGC2019}
R.~Wang, Y.~Gu, H.~Chen, Y.~Li, and B.~Vucetic, ``On the age of information of
  short-packet communications with packet management,'' in \emph{Proc. IEEE
  Global Commun. Conf.}, Waikoloa, HI, Dec. 2019, pp. 1--6.

\bibitem{Cao2021AgeDelay}
J.~{Cao}, X.~{Zhu}, Y.~{Jiang}, Z.~{Wei}, and S.~{Sun}, ``Information age-delay
  correlation and optimization with finite block length,'' \emph{IEEE Trans.
  Commun.}, vol.~69, no.~11, pp. 7236--7250, Nov. 2021.

\bibitem{Polyanskiy2010}
Y.~Polyanskiy, H.~V. Poor, and S.~Verdu, ``Channel coding rate in the finite
  blocklength regime,'' \emph{IEEE Trans. Inf. Theory}, vol.~56, no.~5, pp.
  2307--2359, Apr. 2010.

\end{thebibliography}

\end{document}